# Experimental investigation on the effect of temperature on the frequency limit of GaAs/AlGaAs and AlGaN/GaN 2DEG Hall-effect sensors

Anand V Lalwani[1], Abel John[1], Satish Shetty[5], Miriam Giparakis[2], Kanika Arora[4], Avidesh Maharaj[1], Gottfried Strasser[2], Aaron Maxwell Andrews[2], Helmut Koeck[3], Alan Mantooth[5], Gregory Salamo[5†], and Debbie G Senesky[4*],

†Fellow, IEEE
*Senior Member, IEEE

*Abstract*—This follow-on work investigates the effect of temperature on the frequency limit of 2-dimensional electron gas (2DEG) Hall-effect sensors. Two heterostructure architectures are studied–wideband gap aluminum-gallium-nitride/gallium-nitride (AlGaN/GaN) and high-mobility aluminum-gallium-arsenide/gallium-arsenide (AlGaAs/GaAs). The Hall-effect frequency limit's effect on temperature is further investigated by studying the sheet resistance, sheet density, and mobility of the heterostructures as a function of temperature. Hall-effect sensors are deployed in various environments and often are required to operate at elevated temperatures. Understanding the frequency limit for Hall-effect sensors at higher temperatures and the material effects on temperature range will enable the deployment of sensors for wider applications, such as jet engines. The trade-off between frequency limit and temperature range for the two 2DEG high mobility Hall-effect sensors is explored.

## I. INTRODUCTION

Hall-effect sensors have emerged as crucial components in a wide range of applications, encompassing various fields such as automotive, industrial, medical, and power electronics [1-3]. Named after the American physicist Edwin Hall, who discovered the effect in 1879, Hall effect sensors have revolutionized the way we measure and detect magnetic fields. These sensors utilize the Hall-effect, which describes the generation of a voltage difference perpendicular to an electric current flow in a conductor placed in a magnetic field. The unique properties and capabilities of Hall-effect sensors have led to their extensive utilization in diverse industries, including position sensing, current sensing, speed measurement, proximity detection, and more [4].

However, Hall-effect sensors are susceptible to imperfections in fabrication, piezoelectric effects, and thermal gradients – all of which can lead to offset voltage, which produces noise in sensor readings. Current spinning is a technique used to mitigate or eliminate offset noise in Hall-effect sensors. In current spinning, the direction of the current passing through the sensing element is periodically reversed at a high frequency. This alternating current direction causes the Hall voltage to alternate as well, and the voltage is measured with the remaining two contacts [5]. An average of the measured voltage is taken to negate the offset, yet a minor offset in the microvolt range (μV) may persist.

For all its benefits, current spinning does presents limitations for the frequencies at which Hall-effect sensors can operate. The switching frequency, for instance, should be much higher than that of the magnetic field. If the magnetic field frequency surpasses that of the switching frequency, then current

[1] Department of Electrical Engineering, Stanford University, Stanford, CA 94305, USA
[2] Institute for Solid State Electronics, TU Wien, Gußhausstraße 25-25a, 1040 Vienna, Aust
[3] Infineon Technologies, Villach, Austria
[4] Department of Aeronautics and Aerospace Engineering, Stanford University, Stanford, CA 94305, USA
[5] Department of Electrical Engineering, University of Arkansas, Fayetteville, Arkansas, USA

spinning loses its effectiveness [6]. Current spinning limitations cap the majority of Hall-effect sensors at a frequency limit of 200 kHz [7]. Previous studies demonstrate innovative approaches to do so, such as the X-Hall architecture [6] or the 2-Omega [8] technique, while the authors' previous work investigates the geometry-dependent frequency limit of Hall-effect sensors [9].

Apart from assessing the geometry of Hall-effect sensors, it is necessary to examine their effectiveness in extreme environments [10]. Hall-effect sensors can function at higher frequencies (up to MHz) when using high-mobility 2DEG heterostructures such as GaAs/AlGaAs and AlGaN/GaN. However, the performance of such devices across a range of temperatures has not been investigated. This study builds upon previous research [9] by measuring the frequency limit of high-mobility AlGaN/GaN and GaAs/AlGaAs Hall-effect sensors at elevated temperatures and investigating the influence of material properties on the temperature range where these sensors can perform effectively.

## II. FABRICATION AND EXPERIMENTAL SETUP

The test GaAs/AlGaAs heterostructure on an undoped GaAs substrate was fabricated through molecular beam epitaxy (MBE) technique in a Riber C21 chamber. To begin with, a 50 nm layer of GaAs was deposited as a buffer, followed by a 120 nm GaAs/AlAs superlattice designed to smooth the surface. Si delta doping was used to create the 2DEG layer by adding an average Al percentage of 45% between the 40 nm thick AlAs/AlGaAs SL and the 20 nm thick $Al_{0.33}Ga_{0.67}As$ layer. The design of this two-dimensional electron gas (2DEG) is focused on achieving stability across the whole temperature range, rather than maximizing peak mobility. For further fabrication details, please refer to our previous work [9].

The AlGaN/GaN heterostructure was grown using a nitrogen plasma-assisted Veeco Gen II molecular beam epitaxy (MBE) system. The HEMT structure consists of a 1.5 um GaN buffer, 20 nm $Al_{0.2}Ga_{0.8}N$ barrier, and a 5 nm GaN cap layer, which were deposited on top of a sapphire substrate containing hydride vapor phase epitaxy (HVPE) grown 5 um thick, Fe doped, semi-insulating (SI), GaN template. During the growth of the aforementioned HEMT heterostructure, the substrate temperature was maintained at 796°C. Active nitrogen was supplied by radio frequency plasma with power maintained at 350 W and a flow rate of 0.50 sccm, which corresponded to a 0.25 mL/sec growth rate, with the flux of Ga/N ratio is greater than 1. After completion of the growth process, the excess Ga metal droplets present at the sample surface were removed by rinsing in a Hydrochloric Acid bath for about 20 mins. To process the Greek-Cross Hall-effect sensor device and the Transmission Line Measurement (TLM) structures, we employed a Cl-based Inductively Coupled Plasma (ICP) dry etching process. Further Ohmic contact to the Hall-effect sensor device are made by depositing a Ti (25 nm)/Al (100 nm)/Ni (50 nm)/Au (150 nm) multilayer metal stack, followed by Rapid Thermal Annealing (RTA) at 800°C for 30 s, in 5 sccm flow of $N_2$ atmosphere. For Hall signal and contact resistance characterization purpose, the above fabricated Hall-effect sensor and TLM structure were then mounted on a non-magnetic chip carrier and followed by a wire bonding process.

The characterization setup, shown in Fig. 1(a), utilizes a Lakeshore 8404 Hall measurement tool. The instrument works by applying a DC magnetic field to the sensor and measuring the resulting voltage signal generated by the Hall-effect.

To measure the frequency limit, a square pulse voltage was applied to the Hall-effect sensor under the presence of a magnetic field. The voltage across the opposite terminals of the sensor was measured and recorded on a Keysight InfiniVision DSOX4022A oscilloscope. For further details, please refer to Anand et al. [9]. The primary difference between that investigation and this work is that the Hall-effect sensor was elevated to higher temperatures either in a Lakeshore, as in the case with AlGaN/GaN or using a Linkam chamber and external thermocouples for the AlGaAs/GaAs sensor.

For each sensitivity, frequency, sheet density, and resistance measurement (determined via the Lakeshore), the temperature was modulated from a baseline of 300 K to the maximum respective tolerances of each heterostructure.

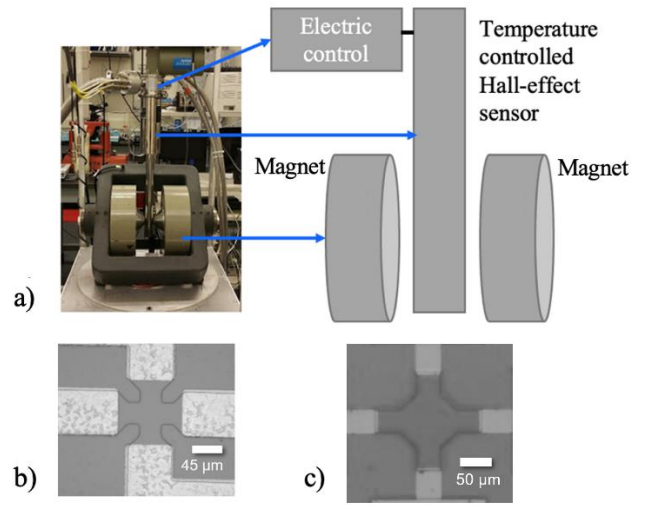

Fig. 1. Experimental setup and sensor designs used for this work. (a) image and block diagram of Lakeshore 8404 Hall measurement tool, (b) is the AlGaN/GaN Hall-effect sensor and (c) is the AlGaAs/GaAs Hall-effect sensor.

## III. RESULTS AND DISCUSSION

Using the defined experimental setup, the frequency limit is evaluated by varying the temperature conditions under which the GaAs/AlGaAs and AlGaN/GaN sensors operate. Further study is done to understand the relationship between the temperature and the current scaled sensitivity, the sheet resistance, and the sheet density of the devices. Fig. 2 shows the results of our investigation with the AlGaN/GaN sensor, while Fig. 3 investigates the GaAs/AlGaAs sensors.

Characteristically, current-scaled sensitivity describes the change in voltage output given a change in magnetic field reading, scaled by the current value. In particular, sensitivity is an important metric when determining the accuracy and precision of a sensor. Comparing Fig. 2(a) (AlGaAs/GaAs) to Fig. 3(a) (AlGaN/GaN), we can see that the AlGaAs/GaAs

Hall-effect sensor sensitivity is significantly higher vs AlGaN/GaN at the same temperature readings. Note that AlGaN/GaN has a larger temperature range than AlGaAs/GaAs throughout our results due to the wider bandgap of AlGaN/GaN [9, 11].

Fig. 2(b) and Fig. 3(b) investigate sheet resistance and frequency as a function of temperature. In both cases, there is a clear inverse relationship between sheet resistance and frequency. This relationship can be attributed to the skin effect, which causes the current to concentrate on the surface of the conductor at high frequencies [4, 12].

Further, Fig. 2(c) and Fig. 3(c) investigates electron sheet density and electron mobility as a function of temperature. It is evident that sheet density and mobility are inversely proportional with respect to temperature. As temperature increases, sheet density increases; meanwhile as sheet density increases, mobility reduces. It follows that mobility is directly proportional to the sheet resistance of the sensor, and sheet resistance is inversely proportional to the frequency limit of the sensor [9]. Evidently, this relationship aligns with the formula (Eq. (1)) for carrier mobility and sheet resistance, thus increasing material sheet resistance inversely impacts the frequency [12].

$$\mu_m = \frac{1}{qn_sR_s} \quad (1)$$

where $q$ is the electron charge, $n_s$ is the sheet density, $R_s$ is the sheet resistance, and $\mu_m$ is the electron carrier mobility.

Moreover, the sensitivity of voltage-biased and current-biased Hall-effect sensors as a function of temperature are investigated in Fig. 2(d) and Fig. 3(d). Previous investigations have shown that current biased sensors should be used when concerned with the temperature stability of a system [11]. That reasoning aligns with our results for the AlGaN/GaN Hall-effect sensor as shown in Fig. 2(c).

and AlGaN/GaN Hall-effect sensors. Since at higher frequencies an increasingly thin surface layer is used to conduct current (referred to as the skin effect), it is necessary to compare the mobility and frequency of our sensors when determining possible usage bottlenecks. AlGaAs/GaAs is observed to have a much higher electron mobility, while the range of temperatures for its use is more constrained than the AlGaN/GaN sensors. While the electron mobility and frequency limit of the AlGaN/GaN sensor also depreciate as the temperature increases; however, its rate of decline is not as extensive as AlGaAs/GaAs, implying that it is capable of operation at temperatures approaching 650 K.

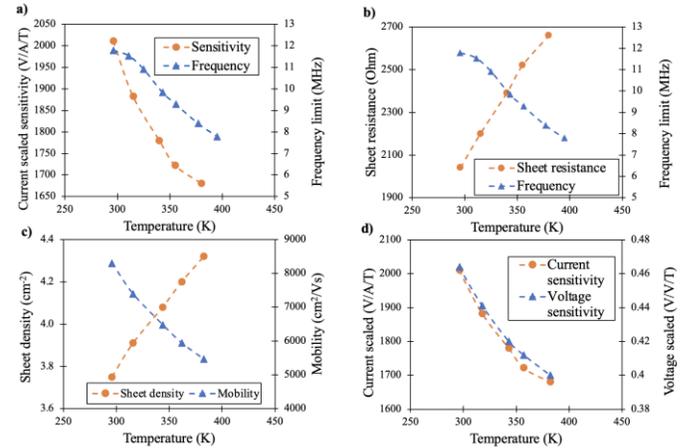

Fig. 3. The experimental investigation of AlGaAs/GaAs Hall-effect sensor sensitivity, frequency limit, sheet density and mobility vs temperature. (a) Investigation of the current scaled sensitivity and frequency as a function of temperature. (b) Investigation of sheet resistance and frequency limit as a function of temperature. (c) Investigation of electron sheet density and electron mobility as a function of temperature. (d) Investigation of voltage biased versus current biased sensitivity as a function of temperature.

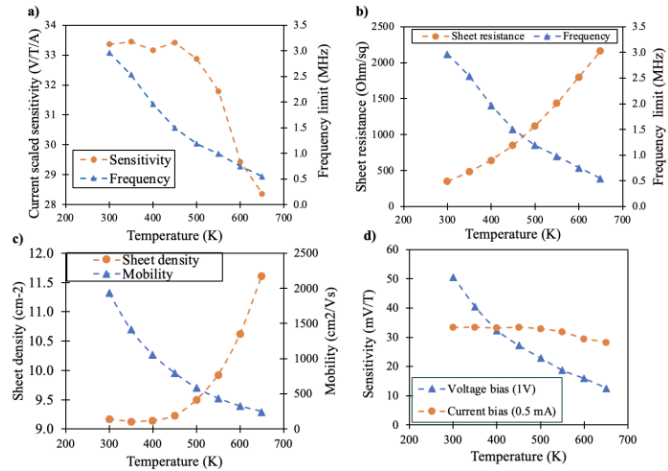

Fig. 2. The experimental investigation of AlGaN/GaN Hall-effect sensor sensitivity, frequency limit, sheet density and mobility with respect to temperature. (a) Investigation of the current scaled sensitivity and frequency as a function of temperature. (b) Investigation of sheet resistance and frequency limit as a function of temperature. (c) Investigation of electron sheet density and electron mobility as a function of temperature. (d) Investigation of voltage biased versus current biased sensitivity as a function of temperature.

Fig. 4 displays the mobility and frequency measurements with respect to temperature, scaled for both the AlGaAs/GaAs

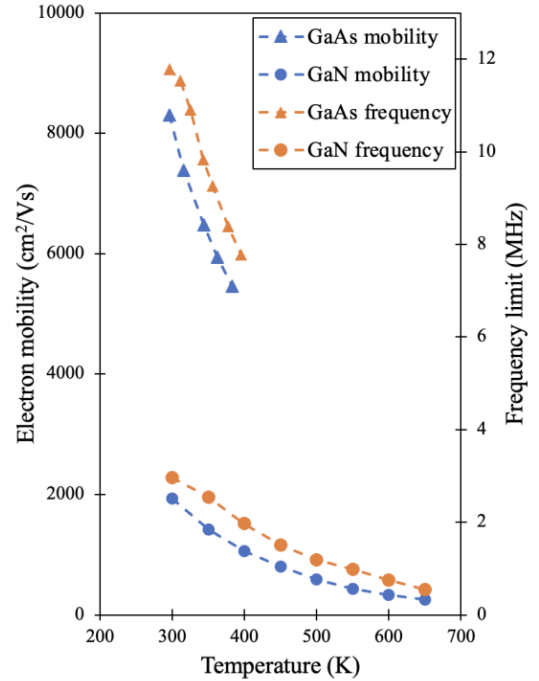

Fig.4. Experimental investigation of electron mobility and frequency limit as a function of temperature for AlGaAs/GaAs and AlGaN/GaN Hall-effect sensors.

Page 3 of 4

Noticeably, there is a clear trade-off between the properties of AlGaAs/GaAs and AlGaN/GaN Hall-effect sensors as the temperature increases. While AlGaN/GaN is capable of operating at higher temperatures, important characteristics such as the frequency limit, sensitivity, and electron mobility reduces as the temperature increases. Conversely, it is also important to note that even AlGaAs/GaAs captures higher frequencies and has a greater mobility than AlGaN/GaN, it remains limited to a smaller temperature range. Rationale for the behavior comes from the bandgaps and mobility (see Table I). AlGaAs/GaAs has much higher mobility, thus needs much lower resistance and be able to operate at higher frequencies. However, AlGaAs/GaAs has a much lower band gap, and thus, as temperatures increase, it sees a substantial rise in sheet density, and the semiconductor becomes extrinsic. AlGaN/GaN, due to its wide bandgap, can withstand higher temperature operations over AlGaAs/GaAs and silicon; however, due to lower mobility intrinsic to the material fabrication, the overall frequency range of AlGaN/GaN is limited. These results implies that AlGaAs/GaAs Hall-effect sensors should be used in high frequency, room-to-lower temperature environments, on the other hand AlGaN/GaN sensors could find applications in lower frequency, higher temperature use cases. Recent advancements in readout circuitry allow Hall-effect sensors to operate higher than 25 MHz [19] and thus optimizing sensor design and material choice can enable applications for such sensors.

TABLE I
Comparisons between AlGaAs/GaAs and AlGaN/GaN Hall-effect sensors over various compiled metrics. Where * symbolizes this work.

| Parameter | Silicon | AlGaN/GaN | AlGaAs/GaAs |
| --- | --- | --- | --- |
| Temperature (°K) | 200[13] | 576[14] | 200[13] |
| Frequency limit (MHz) @RT | 6.3[15] | 2.97* | 46.1* |
| Sensitivity $_{current}$ (V/A/T) | 300[14] | 90[14] | 1881* |
| Sensitivity $_{voltage}$ (V/V/T) | 0.06[14] | 0.057[14] | 0.57* |
| Offset (µT) | 3[14] | 0.5[14] | 35.7* |
| Mobility (cm$^2$/V/s) | 100-400[16] | 1000-2000[17] | 8400-8600[18] |
| Bandgap (eV) | 1.12 | 3.4-6.2 | 1.42-2.16 |

## IV. CONCLUSION

In this work, we establish the effect of temperature on the frequency limit for both GaAs/AlGaAs and AlGaN/GaN Hall-effect sensors. Understanding the temperature's effect on the frequency limit, as well as the other important properties (frequency limit, sensitivity, and electron mobility) studied in this paper, give us a sense of the trade-offs in usage of the two heterostructures for various applications. With the increased development of high-frequency power electronics and the variability of environments in which they are used, it is important to understand how temperature conditions will affect Hall-effect sensor performance. High-mobility devices such as 2DEG-based GaAs/AlGaAs devices allow for higher frequency limits at lower temperatures. Meanwhile, the 2DEG-based GaN/AlGaN system offers significant stability to use cases at extensive high-temperature ranges; although, at the expense of the frequency limit and mobility.